# On the Impact of Aspect Ratio and Other Geometric Effects on the Stability of Rectangular Thermosiphons

**Tri Nguyen**
Ken and Mary Alice Lindquist Department of Nuclear Engineering
The Pennsylvania State University, 228 Hallowell, University Park, PA 16802, USA
nguyen.tri@psu.edu, nguyen.tri@yandex.ru
ASME Member

**Elia Merzari**
Ken and Mary Alice Lindquist Department of Nuclear Engineering
The Pennsylvania State University, 228 Hallowell, University Park, PA 16802, USA
ebm5351@psu.edu, merzari81@gmail.com
ASME Member

**ABSTRACT**

*Single-phase natural circulation thermosiphon loops have been attracting increased interest as they represent the prototype of passive safety systems. However, the stability properties of thermosiphon loops, which can affect and compromise their functionality, are still actively investigated.*

*Traditionally, the stability analysis of thermosiphon loops has been simplified to one-dimensional (1D) calculations, on the argument that the flow would be mono-dimensional when the diameter of the pipe D is orders of magnitude smaller than the length of the loop $L_t$. However, at lower $L_t/D$ ratios, rectangular thermosiphon loops show that the flow presents 3D effect, which also has been confirmed by stability analyses in toroidal loops.*

*In this paper, we performed a series of high-fidelity simulations using the spectral element code Nek5000 to investigate the stability behavior of the flow in rectangular thermosiphon loops. A wide range of $L_t/D$ ratio from 10 to 200 has been considered and the results show many different outcomes compared to previous 1D analytical calculations or stability theory.*

*Moreover, we analyzed the flow in rectangular thermosiphon loops using Proper Orthogonal Decomposition (POD) and we observed that the cases without flow reversal are characterized by swirl modes typical of bent pipes and high-frequency oscillation of the related time coefficients obtained by Galerkin projection. However, the swirl mode was not observed in cases with flow reversals, these cases are characterized by symmetric flow field at 2nd POD mode and the similarity of low-frequency oscillation in the projection of POD modes.*

On the Impact of Aspect Ratio and Other Geometric Effects on the Stability of Rectangular Thermosiphons

1.  INTRODUCTION

Natural circulation systems do not depend on any active power sources, they rely only on convective flow induced by density gradients due to the buoyancy driven force. Single-phase natural circulation thermosiphon loops are examples of natural circulation system which are popularly applied in engineering. In nuclear engineering, thermosiphons have been attracting increased interest as they represent the prototype of passive safety systems, especially for Gen 3+ and Gen 4 reactors. However, oscillations of velocity, pressure and temperature fields can affect thermosiphons and compromise their functionality. Consequently, in order to enhance the proper operation of thermosiphons, the stability dynamics of single-phase free convection need to be carefully investigated.

Single-phase natural circulation thermosiphon loops are characterized by a known set of instabilities that occur for a certain combination of the operational parameters [1]. Nayak et al. [2] investigated a square loop which is heated at the bottom and cooled at the top. Because of the symmetric nature of the problem, two symmetric, convective steady-state solutions exist for each combination of the Grashof and Prandtl numbers, corresponding to a clockwise and a counterclockwise flow. Such solutions might become unstable, leading to a Hopf bifurcation [3, 4] and to the establishment of flow pulsations. Figure 1 provides a visual description of a typical flow reversal, using Computational Fluid Dynamics.

The presence of such instabilities has been extensively investigated with reduced order mono-dimensional (1D) models. We note that this assumption is valid with the consideration that the diameter of the pipe D is orders of magnitude smaller than the length of the loop $L_t$. The first comprehensive study of 1D thermosiphon loop is due to Welander (1967) [5]. Welander studied a loop consisting of a point heat source and sink with imposed wall temperature and heat transfer coefficient. Source and sink are connected by two adiabatic legs along which the working fluid is carried by buoyancy-driven force. Welander found out that the unstable behaviors of the loop are associated with thermal anomalies in the fluid that are advected materially around the loop. P.K. Vijayan and H. Austregesilo [6] developed 1D dimensionless balance equations for a rectangular thermosiphon loop in which the length of the loop over diameter ratio $L_t/D$, modified Grashof and Stanton numbers are the key parameters to determine the stability of the loop. They used 1D linear stability analysis to build stability maps for a limited number of $L_t/D$ in which the $L_t/D$ is greater than 100. L. Cammarata et.al [7] have performed stability analyses of a rectangular thermosiphon loop through the computation of the complex eigenvalues of the Jacobian matrix of the 1D linearized model. They built stability maps for all range of geometrical configuration of the loop, expressed by $L_t/D$ ratio as a function of modified Grashof numbers. We note that of these studies predict that an increase in $L_t/D$, leads to increasing stable behavior due to the stabilizing effect of friction.

Sano [8], however showed that, at very low $L_t/D$ ratio, the flow field often contains 3D features that cannot be neglected. Hence, a fully 3D CFD approach should be applied to correctly predict the flow behavior in this geometry. Merzari et al. [9] had performed first of a kind Large Eddy Simulation (LES) and Direct Numerical Simulation of a toroidal thermosiphon and similar geometries, but that work had no modern experimental data to rely on for the validation of flow reversals as the experiments of Sano did not show such behavior.

Recently experiments with a single-phase rectangular loop have been conducted at the L2 facility in Genoa, Italy (Luzzi et al. [10] and Misale [11]). Misale [11] analyzed the fluid temperature at difference input powers of the L2 loop heater. He discovered that the thermal inertia of the loop can play an important role in the loop stability and the present of pipes cannot be ignored. Luzzi [10] also simulated the loop using both CFD and 1D analysis and compared with the experimental data. The work featured a comparison of different turbulence models, which exhibited widespread predictions of the behavior of the L2 loop. Devia and Misale [12] also performed a CFD analysis of the loop. Their simulation could predict the oscillation amplitude and period of temperature difference between heating and cooling sections, but it underestimated the number of oscillations occurring between two consecutive flow reversal.

Following previous work by Merzari et al. [9], we have performed the simulation of the L2 loop using the open-source spectral element code Nek5000 [13] developed at Argonne National Laboratory. In section 3, we confirm that Nek5000 can simulate this class of flows accurately in terms of onset, frequency and

On the Impact of Aspect Ratio and Other Geometric Effects on the Stability of Rectangular Thermosiphons

amplitude of the oscillations [14]. The results give confidence in the use of Nek5000 to investigate stability behavior in rectangular thermosiphons and to confirm that 3D effects play an important role in the stability.

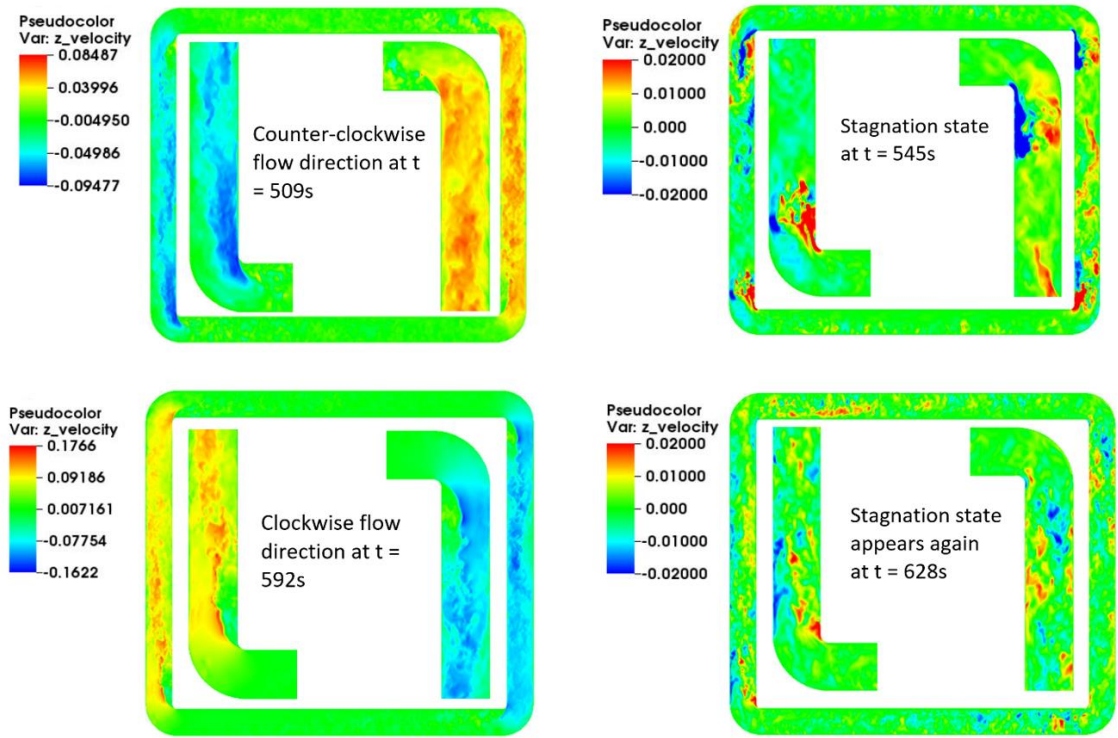

**Figure 1:** An example of flow reversal sequence for case at $L_t/D$ = 50. The flow near elbows (i.e., the top one on the right and the bottom one on the left) are zoomed for better readability.

Thereafter in section 4, we have conducted a series of LES investigating a wider range of the parameter space with a lower $L_t/D$. We confirmed that the stability of the flow strongly depends on the $L_t/D$ ratio [15]. The $L_t/D$ influences the critical Rayleigh number, as well as the frequency of the oscillations. We noted that for $L_t/D \leq 20$ flow reversals do not occur. This cannot be predicted by stability maps derived from 1D analytical analysis from the work of L. Cammarata et.al [7], which shows that all cases with $L_t/D < 117$ and $Gr_m > 2*10^5$ are unstable. In fact, L. Cammarata et.al [7] observes that the stability is lower as $L_t/D$ decreases. However, he noted that the reduction of the system stability is particularly strong for the lower values of the diameter, whereas it becomes almost negligible for the higher diameters. 1D analysis cannot predict a sudden increase in system stability at low $L_t/D$.

Consequently, we postulate that the aspect ratio primarily affects the flow by introducing stronger and more prevalent secondary flows that perturb the overall stability behavior. This leads to the hypothesis that artificially-induced secondary flows can affect the stability behavior of the loop. To verify this hypothesis, we have performed a series of LES for cases at $L_t/D$ = 50 with obstacles which mimic the secondary flows at lower aspect ratios and we could confirm that carefully designed obstacles are effective in changing the overall stability behavior of the thermosiphon. This should not come as a surprise: it is well known that, to stabilize unstable natural circulation, flow stabilizers can be introduced at selected sites [6]. These flow stabilizers can be simply flow restrictions, usually in the form of an orifice. This effect is often attributed to the stabilizing effect of friction and form losses [6]. Misale, M., and Frogheri [16,17] experimentally showed that the localized pressure drops in form of 2 sharp-edged orifices located in the middle of the vertical legs could stabilize the loop behavior.

In section 5, we attempt to provide some additional insights into the physics of flow reversals and the effect of the aspect ratio. The complexity of the intrinsic nature of buoyancy-driven flows makes it difficult to unfold the instability dynamics in rectangular thermosiphons. The mechanism behind the stability behavior of thermosiphons had first been explained by Welander [5]. He noted that the unstable motions of the flow are associated with the thermal anomalies which amplify through the correlated variations in flow rate. Specifically, a warm pocket of fluid passes quicker through the heat sink than through the heat source. Similarly, the heat sink acts more effectively on a cold pocket of fluid. Luzzi et. al. [10] has a more detailed explanation by introducing the formation of a pair of hot and cold plugs. The increase and decrease of the size of the pair while passing the heat sink (cooler) are the mechanism of the instability. However, these approaches are inadequate to explain differences between the cases with unstable behavior (flow reversals) and the cases without flow reversal. In particular, the effect of low aspect ratio is not well explained by these models.

Proper Orthogonal Decomposition (POD) [18,19,20] can be used to identify the dominant flow structure based on energy approach. POD analysis can recognize the fluctuation dynamics of the flow based on the energy associated to each mode, which is well demonstrated in the works of Noorani, A. & Schlatter [21], and Lorenz Hufnagel et al. [22] for bent pipes. In this work, a series of POD analysis for cases with and without flow reversal have been conducted. The objective is to attempt to provide some additional insights into the physics behind the flow reversal phenomena.

## 2. METHODS

In this work we performed a series of LES. We consider the incompressible Navier-Stokes equations of a Newtonian fluid subject to buoyancy, modeled though the Boussinesq approximation:

$$\frac{\partial u_i}{\partial x_i} = 0 \tag{1}$$

$$\frac{\partial u_i}{\partial t} + \frac{\partial}{\partial x_j}(u_i u_j) = -\frac{1}{\rho}\frac{\partial p}{\partial x_i} + \nu \frac{\partial^2 u_i}{\partial x_j \partial x_j} - \beta g_i (T - T_0) \tag{2}$$

$$\frac{\partial T}{\partial t} + u_j \frac{\partial}{\partial x_j}(T) = \frac{\lambda}{\rho c_p}\frac{\partial^2 T}{\partial x_j \partial x_j} \tag{3}$$

where Eq. (1) maintains mass continuity, Eq. (2) maintains momentum continuity, Eq. (3) is the enthalpy equation expressed in terms of temperature. ρ is the density of the fluid, $\nu$ represents the kinematic viscosity, $\beta$ represents the expansion coefficient, $\lambda$ is the conductivity and $c_p$ the heat capacity. Implicit summation applies. In the current work, all properties are considered constant. In the solids, only equation (3) is solved without the convective term. At the solid-fluid interface the temperature is considered continuous.

The equations were solved with the open-source, massively-parallel spectral-element solver Nek5000 [13]. The Spectral Element Method (SEM) features a number of attractive properties for high-fidelity fluid simulations [23]. It is a Galerkin-type method that provides good geometric flexibility similar to finite element methods, but also with minimal numerical dispersion/dissipation and rapid convergence properties typical of spectral methods [13]. Nek5000 has also been shown to be highly scalable [23,24,25], which is crucial for high-fidelity simulations such as those performed in this work.

In SEM, the domain is discretized into E hexahedral elements and represents the solution as a tensor-product of $N^{th}$-order Lagrange polynomials based on Gauss-Lobatto-Legendre (GLL) nodal points. This leads to roughly $E(N+1)^3$ degrees of freedom per scalar field. The discrete Poisson equation for the pressure is solved using a variational multigrid GMRES method with local overlapping Schwarz methods for element-

On the Impact of Aspect Ratio and Other Geometric Effects on the Stability of Rectangular Thermosiphons

based smoothing at resolution N and ≈ N/2, coupled with a global coarse-grid problem based on linear elements [26]. Viscous terms are treated implicitly with second-order backward differentiation, while non-linear terms are treated by a third order extrapolation scheme.

Nek5000 has received extensive validation in numerous references [27,28]. In the present work, sub-grid scale modeling is treated with an explicit filtering approach [29]. Energy is explicitly removed from the high wavenumber flow structures in each element.

## 3. VALIDATION

We performed simulation of the L2 loop using the open-source spectral element code Nek5000 developed at Argonne National Laboratory and compare simulation results against experimental data as well as the 1D analysis and the CFD results of Luzzi et. al. [10]. The primary objective of this stage is to confirm that Nek5000 can simulate this class of flows with sufficient accuracy.

### 3.1 Numerical setup

We simulate several experiments from the L2 facility in Genoa. The computational model for the L2 loop is shown in Fig. 2 (A) and it includes only the fluid domain. The gravitational direction was chosen to be on the Z axis. Dirichlet boundary conditions (BCs) are applied for the coaxial heat exchanger, Neumann BCs are applied for the heater, imposing a constant heat flux.

A hexahedral conformal mesh was generated for the L2 loop. We use Gmsh [30], an open-source 3D finite element mesh generator. The resulting mesh had 80,000 hexahedra in average. A prismatic boundary layer was created to ensure the first grid point to be below y+ = 1 and a sufficient number of grid points to be included in the viscous sublayer at 4$^{th}$ polynomial order. The wall shear stress was estimated based on expected mass flow rate and verified a posteriori. We also investigated mesh convergence by comparing simulation results for the 4$^{th}$, 5$^{th}$ and 6$^{th}$ polynomial order and we observed no differences in prediction of flow reversal phenomena.

The time step in the simulations is set to a maximum of 10$^{-3}$ s (CFL<0.5) and the gravitational direction was chosen to be on the z axis, as shown in Fig. 2 (A). The explicit filtering for the subgrid-scale model is set to 2% of the highest wavenumber.

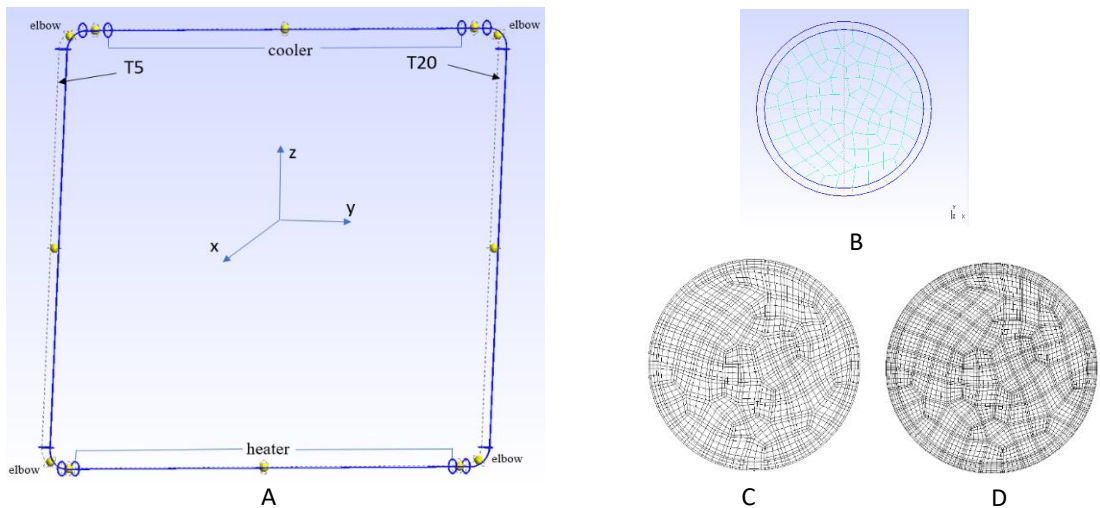

**Figure 2:** A - L2 loop computational model. B - pipe's cross section of the loop mesh. C and D - pipe's cross section of L2 loop mesh for 4$^{th}$ (C) and 6$^{th}$ (D) polynomial order.

On the Impact of Aspect Ratio and Other Geometric Effects on the Stability of Rectangular Thermosiphons

The following boundary conditions are applied at the cooler and heater sections, respectively

$$T = T_c \tag{4}$$

$$k_f \frac{\partial T}{\partial n} = \frac{W}{S} \tag{5}$$

At this stage, W is set at 2000 Watts for all simulation cases, and $T_c$ is adjusted in accordance with the cooler temperature of each simulation case as shown in table 1. The case numbering follows Luzzi et al. [10].

| Case number | Cooler temperature $T_c$(°C) |
|---|---|
| 1 | 4 |
| 1a | 4 |
| 2 | 5 |
| 9 | 18 |
| 9a | 18 |

**Table 1:** The simulation cases and their corresponding cooler temperature.

Case 1 and case 1a refer to the same experimental conditions, but case 1a and 9a are conjugate heat transfer models. The initial conditions for all cases are the same:

$$T = 20\ °C \tag{6}$$

$$u_x = u_y = u_z = 0 \tag{7}$$

The properties of working fluid, correspond to water at 45°C. All simulations are started from a quiescent state. RMS values of the velocity and temperature are evaluated until converged. For all cases, we collect time histories of temperature at coordinates near the inlet and outlet of the cooler for which thermocouple data exists. We also collect data at 9 nearby points, which are on the cross sections.

**3.2 Results and discussion**

The first reversal flow of all cases arrives within the first 100s except in case 1a and 9a, for which the first reversal flow appears at 140s. From that point forward, the behavior of case 1a is similar to case 1. Case 9 and 9a will be discussed separately. This phenomenon is anticipated because case 1a has a solid wall so the system has an increased heat capacity. An example of reversal flow and instantaneous streamwise velocity field are shown in Fig. 3 for case 9. The Z velocity amplitude ($v_z$) of all cases ranges from -0.25 to 0.25. For case 1, $v_z$ ranges from -0.18 to 0.19 [14].

In terms of non-dimensional parameters for case 9, we have Ra = $1.16*10^{11}$ and Re ≈ 7000. The Gr number are defined in the works of Vijayan, P.K. et. al. [6] and Walter Borreani et. al [31]:

$$Gr = \frac{D^3 * \rho_f^2 * \beta * g * W * H}{\mu_f^3 * C_{p\_f} * A_f} \tag{8}$$

The Ra number can be computed using (9):

$$Ra = Gr * \Pr = \frac{4D * \rho_f^2 * \beta * g * W * H}{\mu_f^2 * k_f * \pi} \tag{9}$$

On the Impact of Aspect Ratio and Other Geometric Effects on the Stability of Rectangular Thermosiphons

In table 2, the average oscillation amplitudes are reported by using Fourier analysis [33]. The average amplitude is computed by (10):

$$A = \sqrt{2\frac{\int_{t_0}^{t_1}(\Delta T(t))^2 dt}{(t_1 - t_0)}} \quad (10)$$

We compare Nek5000 results with an OpenFOAM (RANS) model and experimental data from Luzzi et. al. The comparisons are summarized in table 2. The cases have been chosen to reproduce different dynamic conditions.

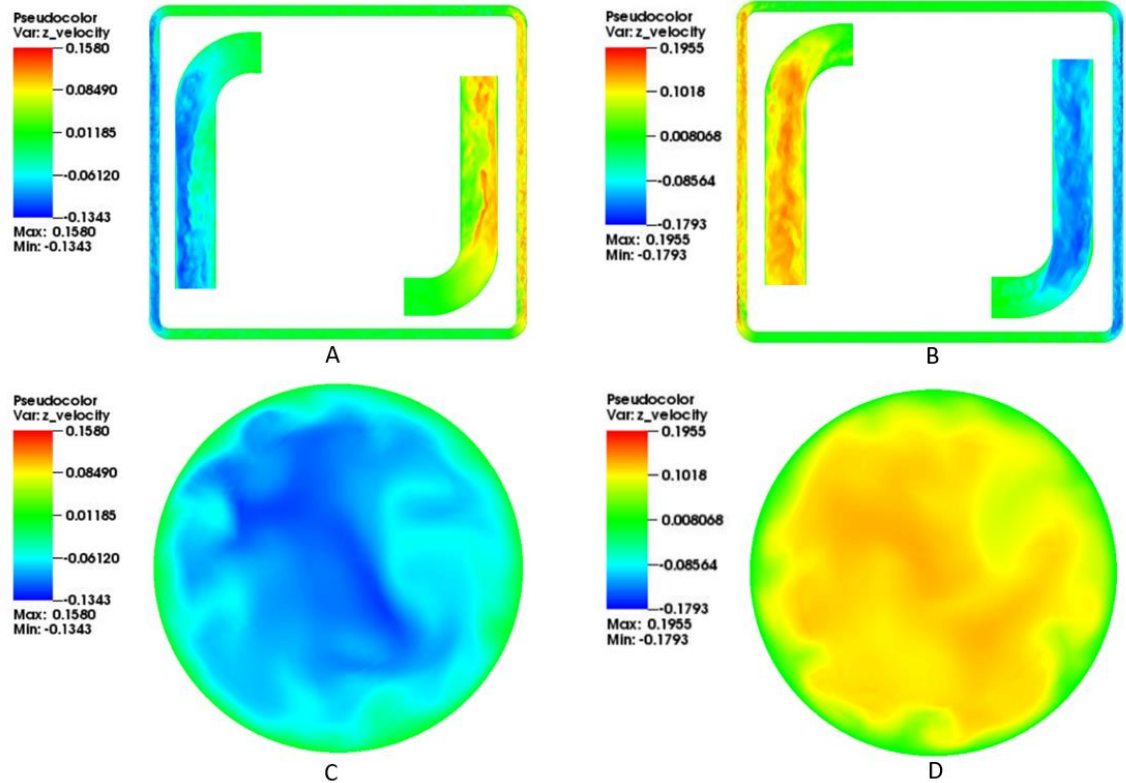

**Figure 3:** A and B – X-plane velocity in the z direction of case 9 at t = 205s (A) and t = 301s (B). The flow near elbows is zoomed for better readability. C and D – Z-plane velocity field of case 9 on the middle of the left leg at t = 205s (C) and t = 301s (D).

Depending on the value of the non-dimensional parameters, flow reversals may or may not occur in rectangular loops [32]. In case 1 as shown in Fig. 4 (A), both Nek5000 and OpenFOAM models (assuming constant wall temperature of the cooler) could reproduce the unstable behavior and the amplitude. However, the oscillation amplitude of Nek5000 is more precise than the OpenFOAM model. The Nek5000 simulation of case 1a as shown in Fig. 4 (B), based on a conjugate heat transfer model, can reproduce more accurately the overall dynamic behavior of the experiment. This is consistent with what was observed Luzzi et al. and it can be explained by a more accurate representation of the thermal resistance and heat capacity of the cooler.

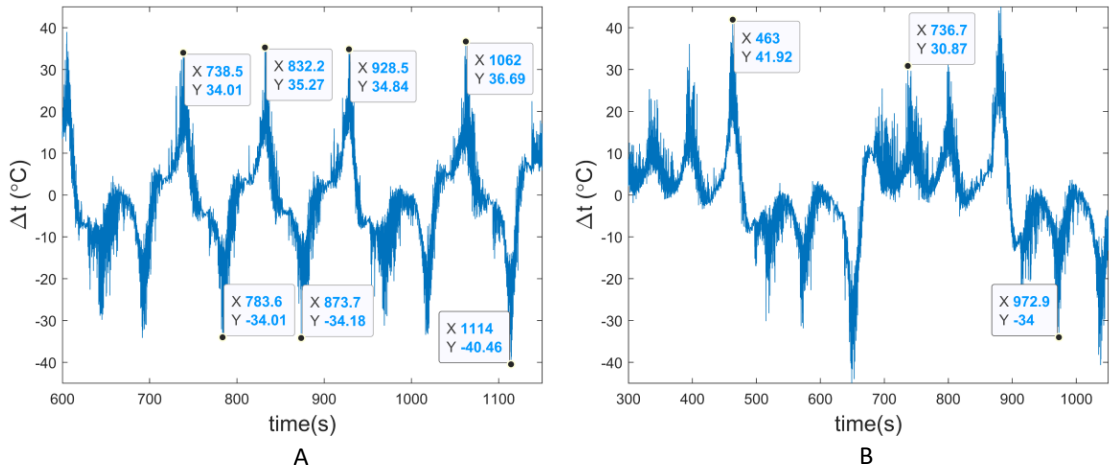

**Figure 4:** A – Nek5000 model of temperature difference across the cooler (δt = t5 - t20) for case 1. B – Nek5000 model of temperature difference across the cooler (δt = t5 - t20) for case 1a.

| | Case 1 | | | | | |
|---|---|---|---|---|---|---|
| | Nek5000 | | Experiment | | OpenFOAM | |
| Amp., $^0$C | 31 | | 33 | | 21 | |
| Period, s | 135 | 97 | 450 | 80 | 100 | |
| | Case 1a (conjugate heat transfer) | | | | | |
| | Nek5000 | | Experiment | | OpenFOAM | |
| Amp., $^0$C | 29 | | 33 | | 10 | |
| Period, s | 430 | 80 | 450 | 80 | 450 | 80 |
| | Case 2 | | | | | |
| | Nek5000 | | Experiment | | 1 D O-O | |
| Amp., $^0$C | 30 | | 30 | | 20 | |
| Period, s | 80 | 145 | 100 | 150 | 120 | 200 |
| | Case 9 | | | | | |
| t ≤ 1500s | Nek5000 | | Experiment | | OpenFOAM | |
| Amp., $^0$C | 27 | | 25 | | 20 | |
| | Case 9a (conjugate heat transfer) | | | | | |
| t ≤ 1500s | Nek5000 | | Experiment | | OpenFOAM | |
| Amp., $^0$C | 27 | | 25 | | 20 | |

**Table 2:** Numerical data on average oscillation amplitude and period of nek5000, experiment [10] and OpenFOAM [10]. Experiment and numerical results from [10] are estimated based on published report.

As shown on Fig. 5 and table 2, case 9 and 9a represent another scenario, where the flow reversals are established but slowly die out and the system reverts to a quasi steady-state. For case 9, the oscillation period in the Nek5000 simulations becomes stable at t = 400s, then the period starts increasing. We note that in case 9, without the explicit modeling of the pipes the flow tends to remain oscillatory, and no conclusion can be reached concerning a quasi steady-state. However, if the solid pipe is explicitly modeled (case 9a) flow reversals stop occurring from t ≈ 1500s, and a quasi-steady-state is observed consistently with the experimental data. Hence, we conclude that the solid domain should be included in this case in order to reproduce the flow dynamic behavior after 1500s. For this case, we note that the OpenFOAM results of Luzzi et al. showed a remarkable dependency on the turbulence model, which is absent here.

At this point, we are confident to conclude that Nek5000 could reproduce remarkably well the dynamic behavior of the system in terms of oscillation amplitude and period of reversals. Moreover, we note that the LES results presented here behaved better or as well as the OpenFOAM RANS results of Luzzi et al. This is anticipated because LES reduces greatly the impact of the turbulence modeling.

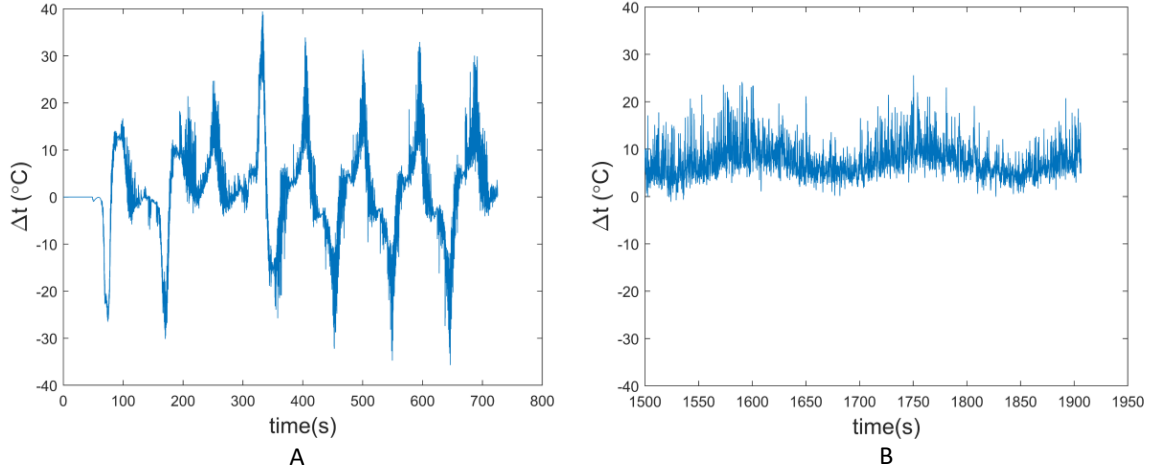

**Figure 5:** A - Nek5000 model of temperature difference across the cooler (δt = t5 - t20) for case 9. B - Nek5000 model of temperature difference across the cooler for case 9a at t > 1500s.

### 3.3 Sensitivity to modeling assumptions and numerical practices

In this section, sensitivity analysis has been performed for the case with $T_c$ = 4 $^0$C (case 1 in Table 2) testing various assumptions.

In particular, the Low-Mach approximation [34] has been applied to address the working fluid's physical properties changes in agreement with the temperature variation in the loop. In Nek5000, the low-Mach-number approximation is achieved by adding a source term to the mass continuity equation (1). The difference between low-Mach Navier-Stokes equations and incompressible Navier-Stokes equations is the source term in (1) which accounts for the change in density. The equations (1), (2), (3) now becomes:

$$\frac{\partial u_i}{\partial x_i} = -\frac{1}{\rho}\left(\frac{\partial \rho}{\partial t} + u_i \frac{\partial \rho}{\partial x_i}\right) \tag{11}$$

$$\frac{\partial}{\partial t}(\rho u_i) + \frac{\partial}{\partial x_j}(\rho u_i u_j) = -\frac{\partial p}{\partial x_i} + \frac{\partial \tau_{ij}}{\partial x_j} + g_i \rho \tag{12}$$

$$\frac{\partial T}{\partial t} + u_j \frac{\partial}{\partial x_j}(T) = \frac{\lambda}{\rho c_p} \frac{\partial^2 T}{\partial x_j \partial x_j} \tag{13}$$

The added source term in mass continuity equation is calculated from energy equation:

$$-\frac{1}{\rho}\left(\frac{\partial \rho}{\partial t} + u_i \frac{\partial \rho}{\partial x_i}\right) = -\frac{1}{\rho}\frac{\partial \rho}{\partial T}\left(\frac{\partial T}{\partial t} + u_j \frac{\partial}{\partial x_j}(T)\right) = \frac{1}{\rho^2}\frac{\partial \rho}{\partial T}\left(\frac{\lambda}{c_p}\frac{\partial^2 T}{\partial x_j \partial x_j}\right) \tag{14}$$

From (14), $\partial \rho / \partial T$ must be calculated for the properties of the working fluid. Details of the implementation of the low-Mach-number formulation are available in [34]. The initial and boundary conditions are the same as for case 1 in table 2. The oscillation amplitudes and periods of cases with different approximations and polynomial orders are shown quantitatively in table 3. As part of this investigation on sensitivity we performed analysis to examine the explicit modeling of the duct walls (conjugate heat transfer) and mesh convergence (5[th] and 6th polynomial order cases). We compared the low-Mach approximation case with the Boussinessq approximation, and no significant differences have been observed. We also verified that the solutions from Nek5000 when running at different polynomial orders yield nearly identical values of oscillation amplitude and period. As shown in table 3, there are no significant differences between cases, with the exception of the inclusion of the solid.

In fact, the solid thermal inertia and conduction can affect the stability of the loop under certain circumstances [35]. Misale et. al. [36] showed that the wall thermal capacity and the axial conductivity can decelerate the fluid. For example, the solid domain should be included to reproduce the quasi steady-state behavior at $T_c$ = 18 $^0$C (case 9). We conclude that Nek5000 can adequately predict the period and amplitude of oscillations as well as the number of oscillations with the Boussinesq approximation and with 4[th] order polynomials.

| Parameters / Cases | N | Amplitude, $^0$C | Period, s Small | Period, s Large |
|---|---|---|---|---|
| Boussinessq approximation without conjugate heat transfer | 4 | 31 | 97 | 135 |
| Boussinessq approximation with conjugate heat transfer | 4 | 29 | 80 | 450 |
| Low-Mach approximation without conjugate heat transfer | 4 | 30 | 97 | 135 |
| Boussinessq approximation without conjugate heat transfer | 5 | 32 | 97 | 135 |
| Boussinessq approximation without conjugate heat transfer | 6 | 31 | 97 | 135 |

**Table 3:** The simulation cases at $T_c$ = 4 $^0$C (case 1) with different approximations and polynomial orders (*N*).

## 4. THE EFFECT OF THE ASPECT RATIO ON THE STABILITY

In this section, we examine the effect of the aspect ratio over on the stability. We intentionally selected cases with $L_t/D$ = 10 and 20 that are predicted to be unstable using 1D analytical stability analysis according to Fig. 3 in the work of L. Cammarata [7]. However, our simulation results of these cases show very different outcomes, which confirmed the role of 3D effects when $L_t/D$ ratio is low. Moreover, we also examine the effect that the addition of obstacles has on the stability. The ultimate objective is to understand how curvature and local geometry features affect the stability.

### 4.1 Numerical setup

We use the same numerical approach as in validation stage, the differences are that all cases run under the temperature of the cooler in formula (4) is set at $T_c$ = 4 $^0$C for all cases. The simulation cases with their corresponding $L_t/D$ and Ra number, as well as flow behavior and oscillation period, are reported in table 4.

To confirm that inducing recirculation patterns may affect flow reversals behavior [15], we have conducted a series of LES for the cases at $L_t/D$ = 50 with added obstacles. We used the same problem setup as for case 9 in table 4. The size of the obstacles included the width along the y-axis and the height along the z-axis [15]. The summary of cases is given in table 5. Note that obstacles are set near the corners of the riser legs of the loop for all cases.

We note that, in accordance with previous literature and to the results provided in section 3, the thermal inertia of pipes can play an important role. This implies that additional non dimensional numbers related to

On the Impact of Aspect Ratio and Other Geometric Effects on the Stability of Rectangular Thermosiphons

the solid should be considered. However, in this section of the manuscript we do not consider this effect as we focus on the impact of the aspect ratio and recirculation regions. In fact, in order to minimize the number of non-dimensional numbers involved in our study, the rest of the analysis presented in this manuscript is performed without conjugate heat transfer.

| Case | $L_t/D$ ratio | Ra number | Flow behavior | Oscillation period, s |
|---|---|---|---|---|
| 1 | 10 | $Ra_{ref}$ | stagnation | |
| 2 | 10 | 100* $Ra_{ref}$ | stagnation | Not present |
| 3 | 10 | $Ra_{ref}/100$ | stagnation | |
| 4 | 20 | $Ra_{ref}/100$ | stagnation | |
| 5 | 20 | $Ra_{ref}/2$ | No flow reversal | |
| 6 | 20 | $Ra_{ref}$ | No flow reversal | Not present |
| 7 | 20 | 7*$Ra_{ref}$ | No flow reversal | |
| 8 | 20 | 100*$Ra_{ref}$ | No flow reversal | |
| 9 | 50 | $Ra_{ref}$ | flow reversal | 160 |
| 10 | 100 | $Ra_{ref}$ | flow reversal | 105 |
| 11 | 200 | $Ra_{ref}$ | flow reversal | 95 |

**Table 4**: The simulation cases with their corresponding $L_t/D$ and Ra number. $Ra_{ref} \approx 10^{11}$.

| Case | Number of obstacles | Size of the obstacle Width | Size of the obstacle Height | Flow reversal |
|---|---|---|---|---|
| 12 | 1 | 0.35D | 0.005$L_t$ | Yes |
| 13 | 1 | 0.85D | 0.005$L_t$ | Yes |
| 14 | 1 | 0.5D | 0.005$L_t$ | Yes |
| 15 | 1 | 0.5D | 0.003$L_t$ | Yes |
| 16 | 2 | 0.5D | 0.005$L_t$ | Yes |
| 17 | 1 | 0.5D | 0.01$L_t$ | Yes |
| 18 | 2 | 0.5D | 0.01$L_t$ | No |
| 19 | 3 | 0.5D | 0.01$L_t$ | Yes |

**Table 5**: The simulation cases at $L_t/D$ = 50 with their corresponding obstacles parameters and their effect to the flow reversal.

**4.2 Results and discussion**

As shown in table 4, the flow reversal happens at $L_t/D \geq 50$ and the oscillation period decreases when the $L_t/D$ increases. For cases 1, 6, 9, 10, 11, the rms of the turbulence fluctuation ($v_z^{rms}$) was calculated for velocity in the z-direction at the same Ra number. The center of the cross section at the middle of the right leg was chosen to measure $v_z$. Assuming the average is $\bar{v}_z$, the rms is computed according to (15). The simulation results demonstrate that, generally, the $v_z^{rms}$ increases as $L_t/D$ increases, as shown in Fig.6 (A). There is an exception for case 1 which has $L_t/D$ = 10. However, we observed that our simulated cases with $L_t/D$ = 10 are always in a stagnation state regardless of Ra numbers, as showed in Fig. 7 (A). We also investigated $L_t/D$ ratio ranging from 10 to 20 to determine the critical aspect ratio above which the stagnation state can disappear. The simulations results show that at $L_t/D$ = 12 and lower, the flow remains in a stagnation state. For $L_t/D$ = 13 and higher, the stagnation state disappears starting at Ra = $Ra_{ref}/2$. Then, a stable flow direction is established. In general, a comprehensive simulation campaign is necessary to determine the exact critical Ra value for each $L_t/D$ ratio, which is beyond the scope of this paper.

On the Impact of Aspect Ratio and Other Geometric Effects on the Stability of Rectangular Thermosiphons

$$v_z^{rms} = \sqrt{\frac{1}{T}\int (v_z - \bar{v}_z)^2 \, dt} \qquad (15)$$

Moreover, we have simulated 5 different cases at $L_t/D$ = 20 (cases 4, 5, 6, 7, 8) and $v_z^{rms}$ is shown in Fig. 6 (B). The simulation results indicate that when Ra/Ra$_{ref}$ = 7, and 100; the stagnation state disappears at t = 140s and then a stable flow direction is established as showed in Fig. 7 (B). We did not observe any flow reversal in these cases.

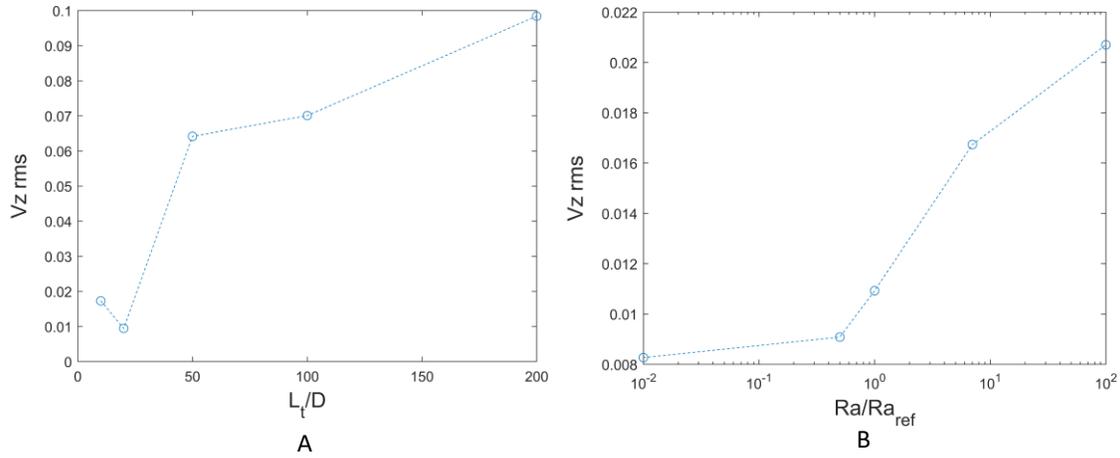

**Figure 6.** A – The turbulence fluctuation strength of velocity in z-direction as a function of $L_t/D$ ratio. B – The turbulence fluctuation strength of velocity in z-direction as a function of Ra/Ra$_{ref}$ when $L_t/D$ = 20.

However, we have observed that the time needed to reach a stable circulation (i.e., time of onset of circulation) starting from a stagnation state depends heavily on the Rayleigh number, as expected. We also noted a correlation between $v_z^{rms}$ and the time of onset of circulation. When Ra/Ra$_{ref}$ = 1, and 0.5; the stagnation state disappears at a later point (t = 200s). For the case with Ra/Ra$_{ref}$ = 0.01, the flow behavior is qualitatively similar to the cases at $L_t/D$ = 10, which means the flow is always in stagnation state (i.e., it is below the onset of convection). We also note that the flow behavior of cases 3, 4, and 5 is similar to the cases (1,2,3,4,7) of Luzzi et. al [10]. but we observe a marked decrease in slope in the relationship between $L_t/D$ and the $v_z^{rms}$. The 3D simulation results of cases from 1 to 8 cannot be predicted by using 1D analytical stability maps of Cammarata et.al [7], which demonstrate that all cases with $L_t/D$ < 117 and Gr$_m$ > 2*10$^5$ are unstable.

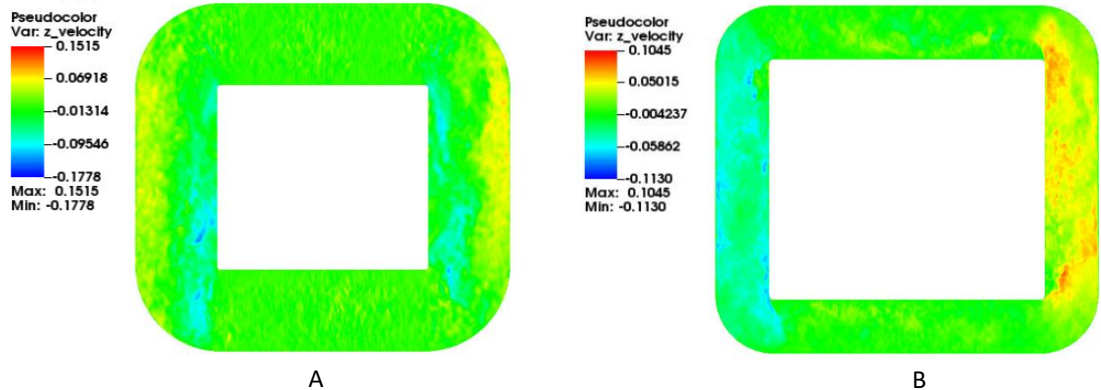

**Figure 7.** A – Stagnation state of cases at $L_t/D$ = 10. B – Stable flow of cases at $L_t/D$ = 20.

For case 12, an increasing in oscillation period has been recorded. Cases 12, 13 and 14 revealed that the oscillation period will increase when the width of the obstacle increases. For case 15, we did not observe any change in oscillation period, as well as the flow reversal behavior compared to case 14. Case 16 with 2 obstacles in an anti-symmetrical pattern and case 17 with an increase in obstacle's height achieved a noticeable increase in oscillation period in comparison with cases 14 and 15, about 1.5x and 1.2x time, respectively. The flow reversal still occurs for all cases from 12 to 17. However, the outcomes of these cases led us to conclude that adding an obstacle will result in an increase of oscillation period and increasing the height of an obstacle over a certain value will also noticeably increase the oscillation period. This confirms our hypothesis that a series of obstacles can alter the stability behavior.

This guided us to case 18: where flow reversals do not occur as shown in Fig. 8. Hence, we conclude that two obstacles can prevent the reversal flow if they are anti-symmetrically placed, and the obstacles have a large enough size. The positions of the obstacles will also determine the direction of the flow. We connect this to potential symmetry-breaking in the heat transfer behavior between clockwise and counter-clockwise circulation for this configuration. We note that the recirculation patterns and form losses in the two directions will be different. However, the stability behavior cannot be simply explained by the mere addition of form and friction losses. In fact, case 19 has a total of three obstacles: two obstacles placed on opposite ends of the same leg and an additional obstacle in the other leg. In this case, we do not observe the same symmetry breaking-effect and reversals do take place, emphasizing the importance of location, number, and size of recirculation regions.

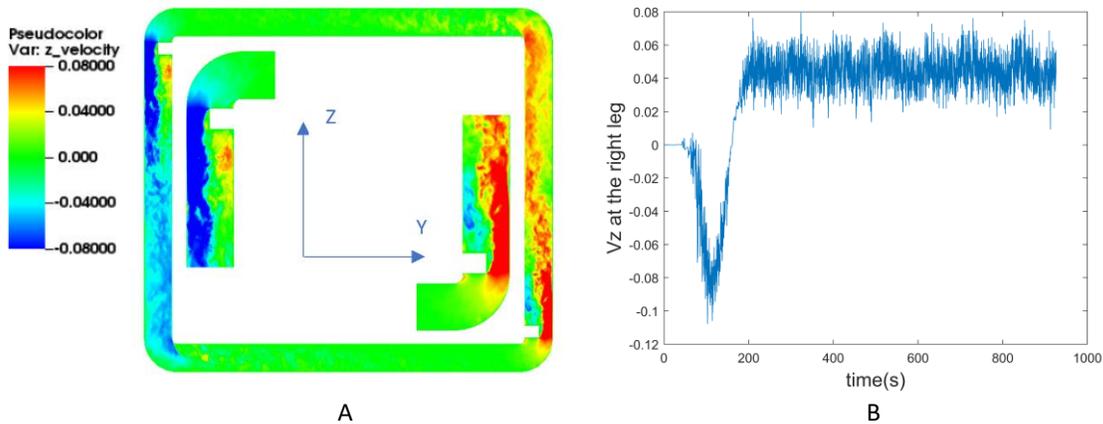

**Figure 8.** A – X-plane velocity in z-direction of case 7 at t = 661s. B – Nek5000 model of $v_z$ at the top leg for case 18.

We conclude that the $L_t/D$ ratio [37] and Ra number have both a strong impact on the stability of the loop. The rms of the streamwise velocity increases with the $L_t/D$ ratio, at the same Ra number. At sufficiently low Ra number, no natural circulation is observed, and that critical Ra number for the onset of convection depends on $L_t/D$. We also confirmed our hypothesis that secondary flows (including artificially induced recirculation) have a profound effect on the dynamic of the loop. These include: increasing/decreasing the oscillation period and potentially stopping the flow reversal. However, we note that the size of the obstacles, their position and quantity also play an important role. In particular, we observe that in our study only two anti-symmetrically placed recirculation regions with large enough size could prevent the flow reversal.

In the next section we will examine in more detail the dynamics of all the cases examined using proper orthogonal decomposition (POD).

**5. POD ANALYSES AND ENERGY PROJECTION**

**5.1 Methods**

The Proper Orthogonal Decomposition (POD) has been introduced to extract a basis for a modal decomposition from a set of snapshots and rank them by their energy [18]. It is a mathematical method to identify the dominant flow structure. In this work we used ''snapshot POD'' [19]. 1200 instantaneous snapshots, sufficiently separated in time to avoid excessive correlation, are taken for each simulation case. Each instantaneous snapshot contains velocity field, and the fluctuating velocity components obtained by subtracting the mean velocity value are arranged in a matrix X. We can write the singular value decomposition (SVD) [19] for matrix X so that:

$$X = U\Sigma V^T \tag{16}$$

$V^T$ is the autocovariance matrix and $VV^T = V^TV = I_{1200*1200}$ where matrix I is the identity matrix. $V^T$ contains ranked reconstructed snapshots and the diagonal matrix $\Sigma$ contains the ranked singular value and the corresponding ranked reconstructed snapshot. Then the eigenvalue decomposition is applied so that:

$$X^TX = V\Sigma^2 V^T \tag{17}$$

Where matrix $X^TX$ is a 1200*1200 square symmetric and positive definite, Matrix $\Sigma^2$ contains ranked eigenvalues. We also performed energy projection for the first 3 energetic modes. At a given time t, the amount of energy contained in a given mode is computed according to (18).

$$E^i(t) = \int R_x^j v_x(x,y,z,t)\,dxdydz + \int R_y^j v_y(x,y,z,t)dxdydz + \int R_z^j v_z(x,y,z,t)dxdydz \tag{18}$$

The superscript i is used to indicate the mode number (i=1,2,3); $R_x^j$, $R_y^j$, $R_z^j$ are the velocity components of eigenvector j (j=1,2,3). Through case 1-19 as shown in table 4 and 5, we picked some cases with and without flow reversal to conduct POD and energy projection. The details are given in table 6.

We use the same numerical approach and problem setup as in section 3 of this work. Noted that all cases in table 4 and 5 have the same curvature radius at $90^0$ bent $R/R_c$ ratio = 0.95. To investigate the effect of the local curvature, the additional $R/R_c$ = 0.5 set of cases have been developed for cases 6 and 9. We found no difference when comparing POD analysis and energy projection results with these cases at $R/R_c$ = 0.95. However, we observed that at the same $L_t/D$ ratio, cases at $R/R_c$ ratio = 0.5 has longer stagnation time before the onset of convection in comparison with cases at $R/R_c$ ratio = 0.95. Once convection is established, the flow becomes stable and the dynamic behavior of these cases is the same.

| Case | $L_t/D$ ratio | $R/R_c$ ratio | Flow reversal |
|---|---|---|---|
| 6 | 20 | 0.5 | No |
|   |    | 0.95 | No |
| 9 | 50 | 0.5 | Yes |
|   |    | 0.95 | Yes |
| 10 | 100 | 0.95 | Yes |
| 17 | 50 with 1 obstacle | 0.95 | Yes |
| 18 | 50 with 2 obstacles | 0.95 | No |

**Table 6**: POD analysis and energy projection cases.

On the Impact of Aspect Ratio and Other Geometric Effects on the Stability of Rectangular Thermosiphons

## 5.2 Results and discussion

In this section, we discuss results of the POD analysis of the flow field with objective of elucidating the dynamics underlying the thermosiphon loops.

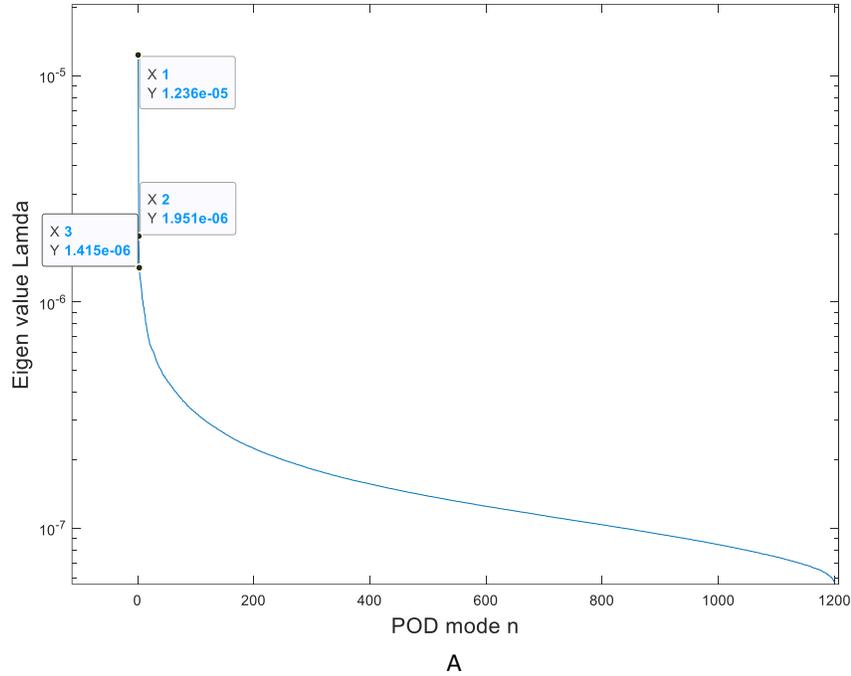

A

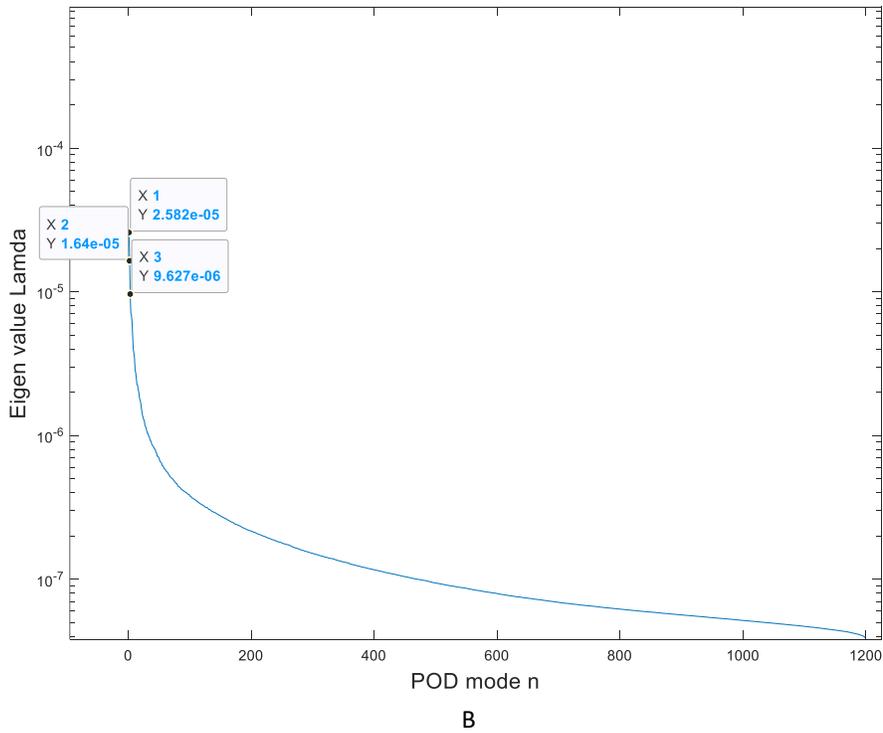

B

**Figure 9**. A – The ranked eigenvalue of case 18 at R/Rc = 0.95. B – The ranked eigenvalue of case 6 at R/Rc = 0.5.

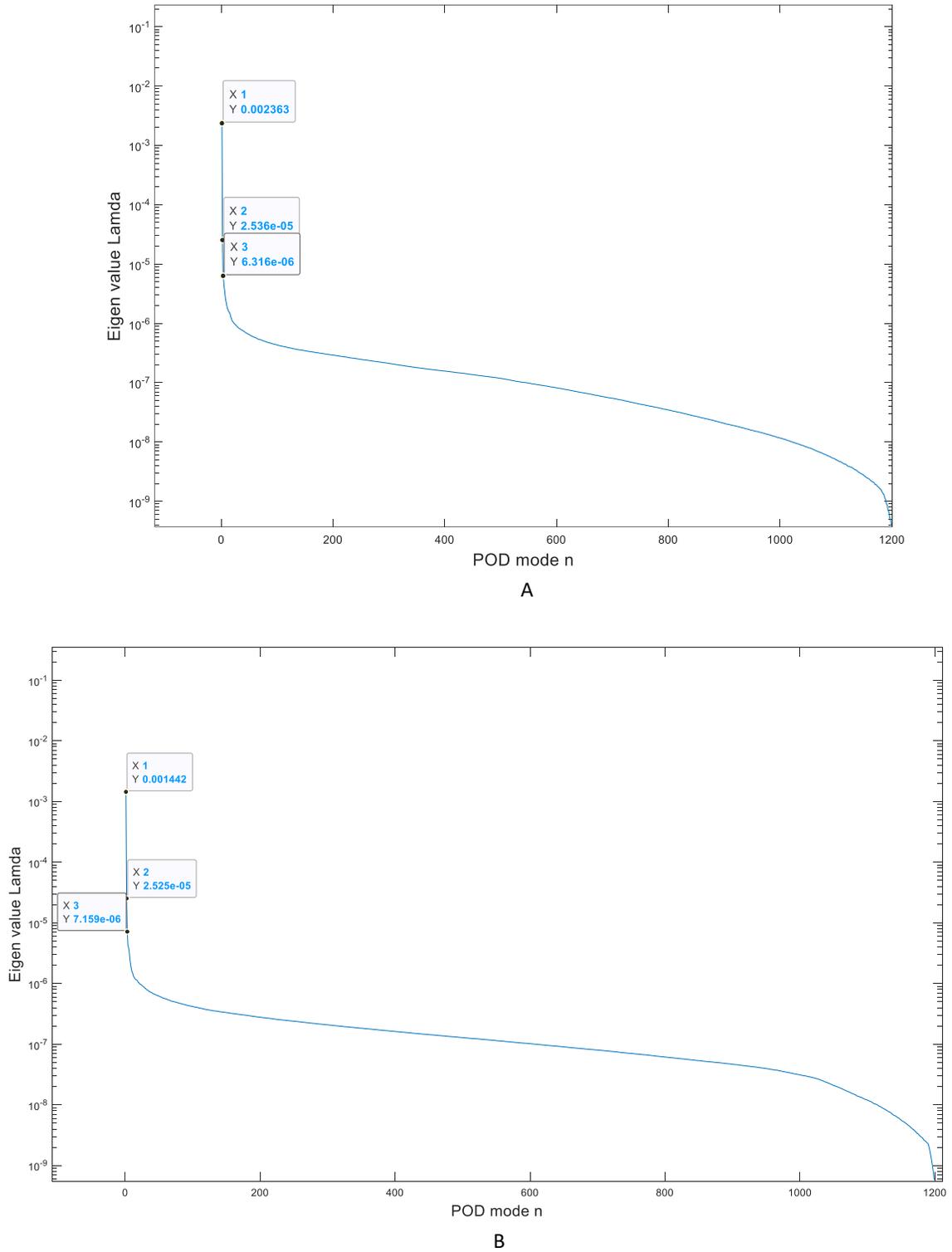

**Figure 10**. A – The ranked eigenvalue of case 9 at R/Rc = 0.95. B – The ranked eigenvalue of case 17 at R/Rc = 0.95.

On the Impact of Aspect Ratio and Other Geometric Effects on the Stability of Rectangular Thermosiphons

Fig. 9 and 10 show the eigenvalue distribution for case 6, 9, 17 and 18. In cases with flow reversals (9, and 17), the 2$^{nd}$ mode has a decrease of 2 orders of magnitude in comparison with the 1$^{st}$ mode, showing a steep energy decrease. This is predictable because the 1$^{st}$ mode is associated with the flow reversal [20], the more unstable the case is, the more energy the 1$^{st}$ mode contains. Cases without flow reversal (6 and 18) have an energy decrease between 1$^{st}$ and 2$^{nd}$ mode within an order of magnitude, showing less eigenvalue dominance. We note that the case at $L_t/D = 20$ (case 6) has the lowest dominance.

Fig.11 displays the comparison of y-plane flow field near 90$^0$ bent of the most energetic mode between cases with and without flow reversal. It is important to note that cases without flow reversal (6 and 18 in table 6) present a strong swirl structure in the 1$^{st}$ and 2$^{nd}$ modes as observed in the work of Noorani, A. & Schlatter [21].

On the other hand, in the cases with flow reversal (9, 10, 17 in table 6) the 1$^{st}$, 2$^{nd}$, and 3$^{rd}$ modes are characterized by Dean vortexes near 90$^o$ bent as observed in the work of Yongmann M. Chung & Zhixin Wang [38]. We did not observe any strong swirl within the first 10 energetic modes. However, the swirl mode becomes dominant in the cases without flow reversal.

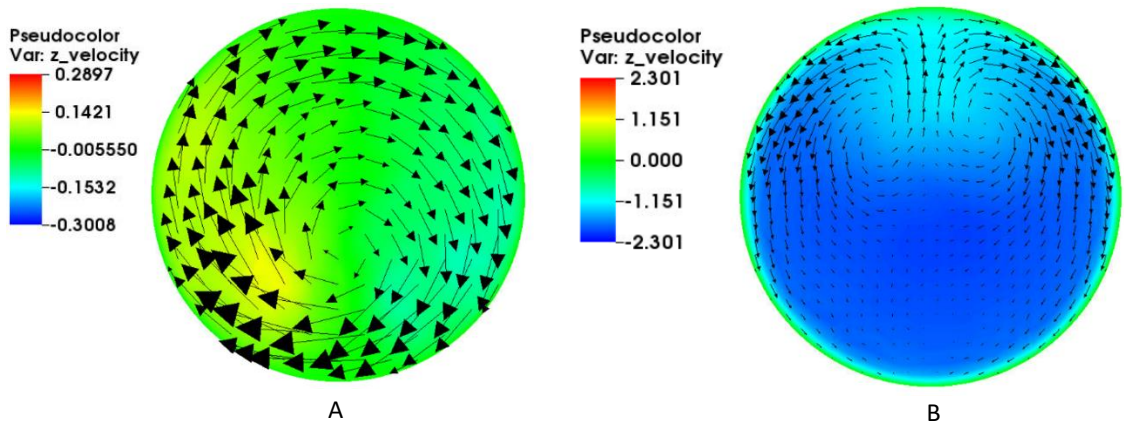

A                                                                B

**Figure 11.** A – Y-plane flow field near 90$^0$ bent of the 1$^{st}$ energetic mode of case 18. B – Y-plane-flow field near 90$^0$ bent of the 1$^{st}$ energetic mode of case 9.

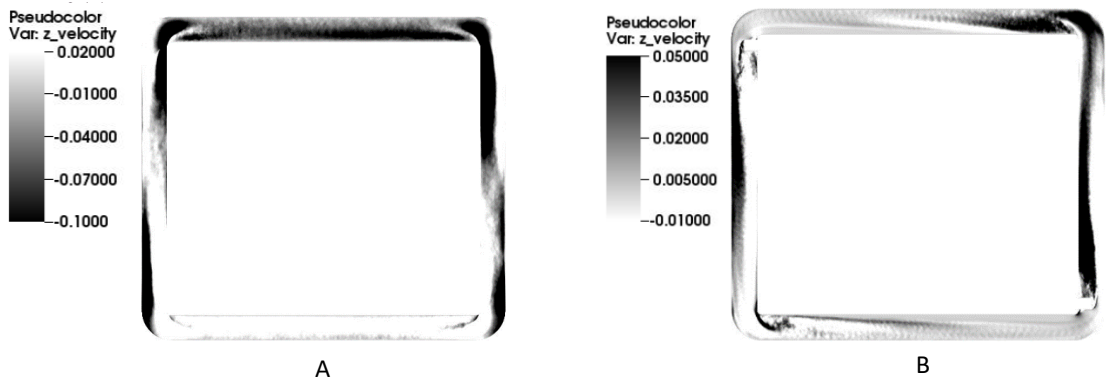

A                                                                B

**Figure 12**. A – X-plane flow field of the 2$^{nd}$ energetic mode of case 9. B – X-plane Flow field of the 2$^{nd}$ energetic mode of case 18.

Let us now take a look on the x-plane flow field of the 2$^{nd}$ and 3$^{rd}$ energetic mode between cases with and without flow reversal. Figure 12 shows the characteristic z-velocity flow field of the 2$^{nd}$ energetic mode of cases with flow reversal (case 9, 10 and 17) and cases without flow reversal (case 6 and 18). The case without flow reversal shows travelling wave type behavior in pair of modes for all 3 most energetic modes. This is consistent with the work of Noorani, A. & Schlatter [21]. However, the cases with flow reversal present a symmetric flow structure in modes 2 and 3. This is unique to cases with flow reversal.

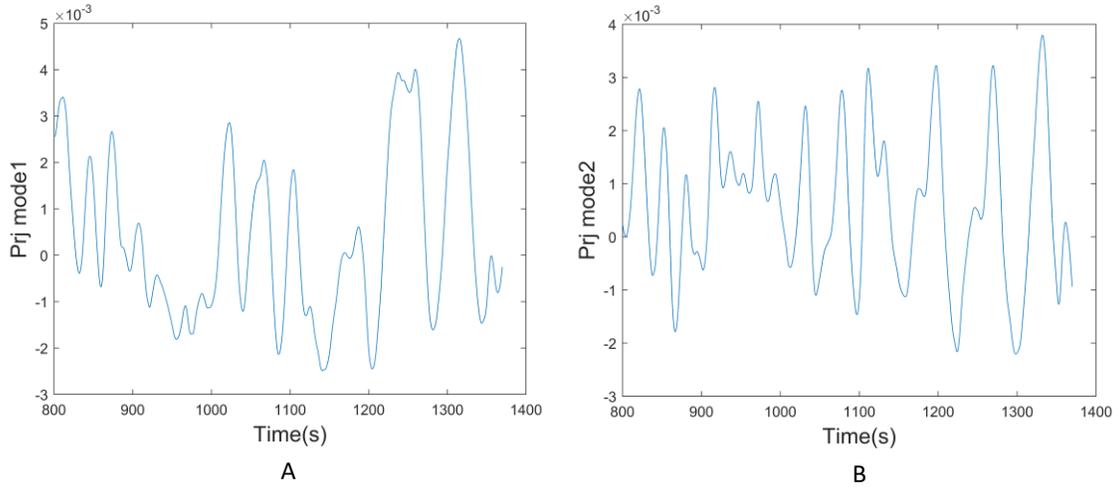

**Figure 13**. Energy projection of the 1st (A) and 2nd (B) modes of case 6.

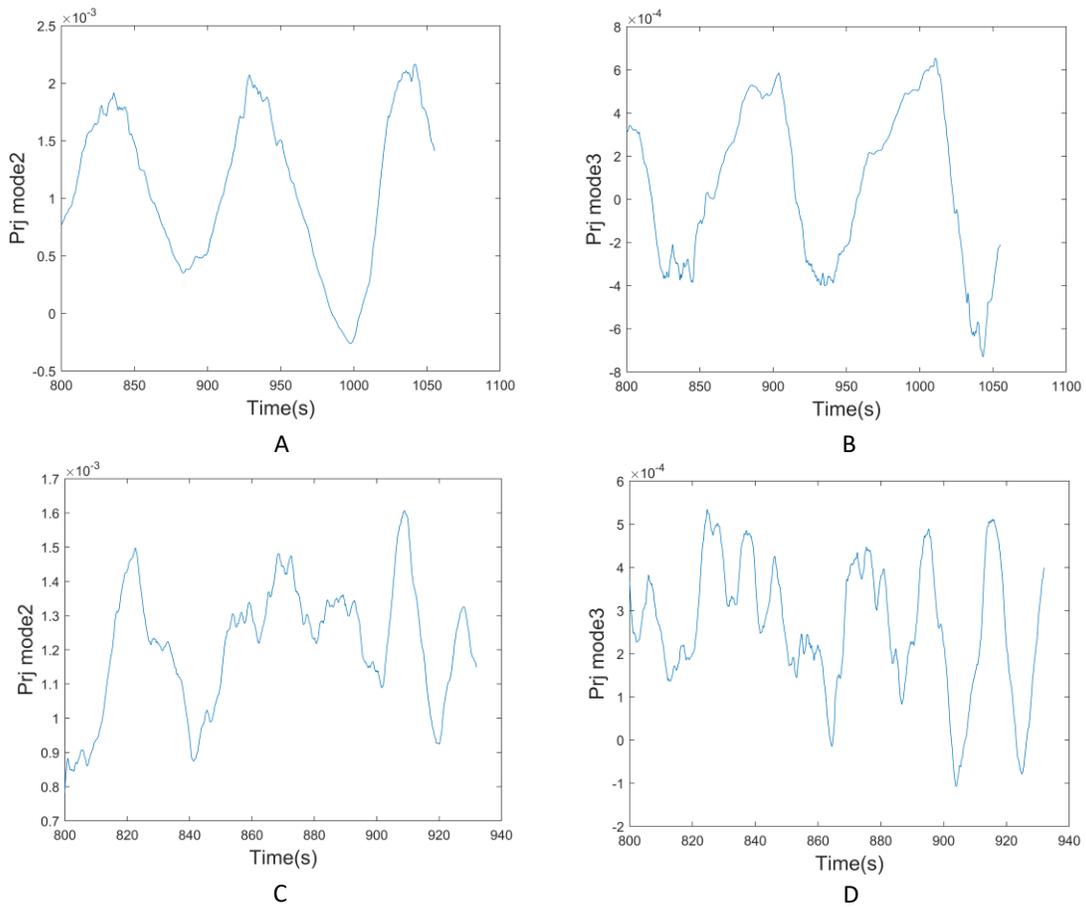

**Figure 14**. A and B – Energy projection of the 2nd (A) and 3rd (B) modes of case 17. C and D – Energy projection of the 2nd (C) and 3rd (D) modes of case 18.

Let us now consider the energy projection. Overall, the energy projection of the 2$^{nd}$ mode of cases without flow reversal has a remarkable higher oscillation frequency than that of 1$^{st}$ mode. An example is shown in Fig. 13 for case 6. The same situation applied for 3$^{rd}$ (D) and 2$^{nd}$ (C) mode as show in Fig. 14 (C and D) for case 18. This is consistent with the dynamic for swirl modes as observed by Noorani, A. & Schlatter [21]. In that work, the 1$^{st}$ mode has a noticeable lower frequency range than the 2$^{nd}$ mode, the same situation applied for 3$^{rd}$ and 2$^{nd}$ mode. The increase in frequency range of power spectral density for time coefficient from the 1$^{st}$ mode to the 3$^{rd}$ mode is consistent with the increase in oscillation frequency in this work.

As shown in Fig. 14 (A and B), the behavior of energy projection of 2$^{nd}$ (A) and 3$^{nd}$ (B) modes of case 17 is similar. The same situation applies for 1$^{st}$ and 2$^{nd}$ modes but these modes are not reported here for the interest of brevity. Generally speaking, the oscillation behavior of energy projections of 1$^{st}$, 2$^{nd}$ and 3$^{rd}$ modes of cases with flow reversal are similar, they have larger oscillation period compared to the one of cases without flow reversal.

In summary, the POD analysis revealed that the dynamics of cases with flow reversal versus cases without flow reversal are radically different, with significant differences in the modes at play. This difference occurs regardless of whether the flow reversal was stopped using a blockage or did not occur naturally due to a lower aspect ratio. Moreover, all cases without flow reversals are dominated by swirl modes. In fact, the case at $L_t/D = 50$ with two obstacles (case 18), shows a remarkable dominance of swirl modes, that were otherwise absent from the most energetic modes before the obstacles were added. This indicates that these modes play an important role in the dynamics of rectangular thermosiphons, a fact that has been somewhat omitted in the literature, to our knowledge.

While this analysis provides some new insights into the dynamics of rectangular thermosiphons, a lot of questions remained unanswered. How does the change in curvature and recirculation inhibit the flow reversal at lower aspect ratio? POD may not be the correct tool to answer this question. As soon as flow reversals occurs, it dominates completely the energy content, hiding perhaps subtle differences in the lower order modes that could help clarify differences in dynamic behavior.

## 6. CONCLUSION

We have performed a series of Large Eddy Simulations of the flow in a single-phase rectangular thermosiphon for various combinations of geometrical parameters and BCs, with the spectral element method.

As a first steps we have confirmed that Nek5000 can simulate this class of flow with sufficient accuracy, which is the basis for continuing investigation the effect of the aspect ratio on the stability of rectangular thermosiphons loops.

We demonstrated that the critical Rayleigh number for the onset of flow reversals depends heavily on $L_t/D$ and both $L_t/D$ ratio and Ra number have a strong impact on the stability of the loop. Moreover, we have confirmed that adding obstacles in specific locations can alter the stability of the loop. In summary we have concluded that that secondary flows induced by curvature or by artificially added obstacles, can affect the oscillation period, even stopping the flow reversal. This seems to confirm the hypothesis raised by Sano [8] that curvature and related re-circulation play an important role in the instability dynamics.

To further investigate the dynamics, POD analyses and energy projection have been conducted for cases with and without flow reversal. We observe that the cases without flow reversal are characterized by swirl modes typical of bent pipes and high frequency oscillation in the time coefficient. To our knowledge, this is the first time this was observed for rectangular thermosiphons. Swirl modes were not observed in cases with flow reversals (high $L_t/D$, without obstacles), suggesting a stronger dominance of bent-pipe physics at low aspect ratios. The detail effect of these modes on the stability will be subject of a future study, employing additional techniques, beyond POD.

On the Impact of Aspect Ratio and Other Geometric Effects on the Stability of Rectangular Thermosiphons**NOMENCLATURE**

| | | |
|---|---|---|
| $L_t$ | | The total length of the loop (m) |
| $D$ | | The diameter of the pipe (m) |
| $R$ | | The radius of the pipe (m) |
| $R_c$ | | The $90^0$ bent curvature radius of the loop (m) |
| $W$ | | Thermal power of the heater (W) |
| $A_f$ | | Flow area (m$^2$) |
| $S$ | | Area of heater surface (m$^2$) |
| $H$ | | The height of the loop (m) |
| $Gr_m$ | | Modified Grashof number |
| $Ra$ | | Raleigh number |
| $Ra_{ref}$ | | Reference Raleigh number |
| $Pr$ | | Prandtl number |
| $\rho_f$ | | Density of working fluid (kg*m$^{-3}$) |
| $\beta$ | | Thermal expansion coefficient of working fluid (1/K) |
| $k_f$ | | The conductivity of working fluid (W*m$^{-1}$*K$^{-1}$) |
| $C_{p\_f}$ | | Isobaric heat capacity of working fluid (J*kg$^{-1}$*K$^{-1}$) |
| $\mu_f$ | | Dynamic viscosity of working fluid (Pa.s) |

On the Impact of Aspect Ratio and Other Geometric Effects on the Stability of Rectangular Thermosiphons

**Figure Captions List**

| | |
|---|---|
| Fig. 1 | An example of flow reversal sequence for case at $L_t/D = 50$. The flow near elbows (i.e. the top one on the right and the bottom one on the left) are zoomed for better readability. |
| Fig. 2 | A - L2 loop computational model. B - pipe's cross section of the loop mesh. C and D - pipe's cross section of L2 loop mesh for $4^{th}$ (C) and $6^{th}$ (D) polynomial order. |
| Fig. 3 | A and B – X-plane velocity in the z direction of case 9 at t = 205s (A) and t = 301s (B). The flow near elbows is zoomed for better readability. C and D – Z-plane velocity field of case 9 on the middle of the left leg at t = 205s (C) and t = 301s (D). |
| Fig. 4 | A – Nek5000 model of temperature difference across the cooler ($\delta t = t5 - t20$) for case 1. B – Nek5000 model of temperature difference across the cooler ($\delta t = t5 - t20$) for case 1a. |
| Fig. 5 | A - Nek5000 model of temperature difference across the cooler ($\delta t = t5 - t20$) for case 9. B - Nek5000 model of temperature difference across the cooler for case 9a at t > 1500s. |
| Fig. 6 | A – The turbulence fluctuation strength of velocity in z-direction as a function of $L_t/D$ ratio. B – The turbulence fluctuation strength of velocity in z-direction as a function of $Ra/Ra_{ref}$ when $L_t/D = 20$. |
| Fig. 7 | A – Stagnation state of cases at $L_t/D = 10$. B – Stable flow of cases at $L_t/D = 20$. |
| Fig. 8 | A – X-plane velocity in z-direction of case 7 at t = 661s. B – Nek5000 model of $v_z$ at the top leg for case 18. |
| Fig. 9 | A – The ranked eigenvalue of case 18 at $R/Rc = 0.95$. B – The ranked eigenvalue of case 6 at $R/Rc = 0.5$. |
| Fig. 10 | A – The ranked eigenvalue of case 9 at $R/Rc = 0.95$. B – The ranked eigenvalue of case 17 at $R/Rc = 0.95$. |
| Fig. 11 | A – Y-plane flow field near $90^0$ bent of the $1^{st}$ energetic mode of case 18. B – Y-plane-flow field near $90^0$ bent of the $1^{st}$ energetic mode of case 9. |
| Fig. 12 | A – X-plane flow field of the $2^{nd}$ energetic mode of case 9. B – X-plane Flow field of the $2^{nd}$ energetic mode of case 18. |
| Fig. 13 | Energy projection of the $1^{st}$ (A) and $2^{nd}$ (B) modes of case 6. |
| Fig. 14 | A and B – Energy projection of the $2^{nd}$ (A) and $3^{rd}$ (B) modes of case 17. C and D – Energy projection of the $2^{nd}$ (C) and $3^{rd}$ (D) modes of case 18. |

**Table Captions List**

Table 1    The simulation cases and their corresponding cooler temperature.

Table 2    Numerical data on average oscillation amplitude and period of nek5000, experiment [10] and OpenFOAM [10]. Experiment and numerical results from [10] are estimated based on published report.

Table 3    The simulation cases at $T_c$ = 4 $^0$C with different approximations and polynomial orders.

Table 4    The simulation cases with their corresponding $L_t/D$ and Ra number. $Ra_{ref} \approx 10^{11}$.

Table 5    The simulation cases at $L_t/D$ = 50 with their corresponding obstacles parameters and their effect to the flow reversal.

Table 6    POD analysis and energy projection cases.